# Phase locking of two independent degenerate coherent anti-Stokes Raman scattering processes

Qun Zhang*

*Hefei National Laboratory for Physical Sciences at the Microscale and Department of Chemical Physics,*

*University of Science and Technology of China, Hefei, Anhui 230026, People's Republic of China*

*qunzh@ustc.edu.cn

We propose an elegant scheme towards phase locking of two independent degenerate coherent anti-Stokes Raman scattering (CARS) processes. The optical implementation involves a modified Mach-Zehnder interferometer that is utilized to transfer phase coherence from three totally uncorrelated laser beams into two degenerate CARS beams which are independently produced in two distinct Raman active samples. Such a phase-transport interferometer allows for explicit measurement and control of phase differences between the two phase-locked degenerate CARS processes, hence may lead to applications in other pertinent research fields.

PACS number(s):
**42.65.Dr** − Stimulated Raman scattering; CARS;
**07.60.Ly** − Interferometers;
**42.25.Hz** − Interference;
**42.25.Kb** − Coherence;
**42.50.Ex** − Optical implementations of quantum information processing and transfer.

As a nonlinear four-wave-mixing process [1-3], coherent anti-Stokes Raman scattering (CARS) [4-7] employs multiple photons ("pump", "Stokes", and "probe") to address the molecular vibrations, and as a result produces a signal in which the emitted light waves are coherent with one another.  The pump and probe fields are usually chosen to be identical, thereby rendering the resultant optical output a so-called coherent anti-Stokes (or CARS) signal. It is well known that the total CARS signal comes from a coherent addition of the signal from individual Raman active molecules.  Although it is also well known that coherence can be generally defined by the correlation properties between quantities of an optical field and that the simplest way to reveal correlations between light waves is making use of interference phenomena [8], an interesting subject pertinent to phase coherence of CARS [9] has not been thoroughly exploited, i.e., is it possible to achieve phase locking between two independent degenerate CARS processes arising from two distinct Raman active media?  In other words, can we obtain the interference pattern from such two independently produced yet energetically degenerate CARS signals?  In this Brief Report, we explore a unique approach and design an optical system to this end, in the hope that it may provide useful insight into the above questions of fundamental interest as well as bring forth potential applications in CARS-related research and in quantum communication based on coherent phase transport.

We start with a general case that involves two independent CARS processes generated from two distinct Raman active samples, (*a*) and (*b*), as schematically depicted in Fig. 1.  Assume that the two samples have their own characteristic Raman modes, $\omega_{fg}^{(i)}$ ($i = a, b$), with *g* and *f* respectively designating the initial ground state and the final (vibrationally) excited state of the samples. We apply a pair of temporally and spatially overlapped laser fields, the pump $\omega_p^{(i)}$ and the Stokes $\omega_S^{(i)}$ ($i = a, b$), on each sample, such that $\omega_{fg}^{(i)} = \omega_p^{(i)} - \omega_S^{(i)}$ ($i = a, b$) are satisfied, i.e., the joint action of the two simultaneously present fields efficiently establish a coupling between the $g$ and $f$ states, making the molecule reside in a coherent superposition of states. We herein confine ourselves to only the situation that the pump field also operates as a probe which simultaneously promotes the system to a virtual state, hence CARS signals with frequencies of $\omega_{CARS}^{(i)} = 2\omega_p^{(i)} - \omega_S^{(i)}$ ($i = a, b$) are expected to emit out of the samples. To observe optical interference between such two CARS signals, one must ensure first their energy degeneracy. To do so, we propose an all-optical scenario as follows. Assume we have three independent laser sources with output frequencies of $\omega_0$, $\omega_1$, and $\omega_2$, respectively. We generate $\omega_p^{(i)}$ ($i = a, b$) by sum-frequency mixing (SFM) $\omega_i$ ($i = 1, 2$) with $\omega_0$ in two SFM nonlinear crystals:

$$\omega_p^{(a)} = \omega_0 + \omega_1, \quad \omega_p^{(b)} = \omega_0 + \omega_2, \qquad (1)$$

and generate $\omega_S^{(i)}$ ($i = a, b$) by difference-frequency mixing (DFM) $\omega_p^{(i)}$ ($i = a, b$) with $\omega_i$ ($i = 2, 1$) in two DFM nonlinear crystals:

$$\omega_S^{(a)} = \omega_p^{(a)} - \omega_2 = \omega_0 + \omega_1 - \omega_2, \quad \omega_S^{(b)} = \omega_p^{(b)} - \omega_1 = \omega_0 + \omega_2 - \omega_1. \qquad (2)$$

Thus we explicitly obtain the following energy relationships:

$$\omega_{CARS}^{(a)} = 2\omega_p^{(a)} - \omega_S^{(a)} \equiv \omega_0 + \omega_1 + \omega_2 \equiv 2\omega_p^{(b)} - \omega_S^{(b)} = \omega_{CARS}^{(b)}, \qquad (3)$$

i.e., the energy degeneracy condition for the two independently produced CARS signals is *automatically* fulfilled through using only the facile nonlinear frequency mixing techniques.

It is worth noting that the above scenario is not only interesting but practically operational considering that many nowadays commercially available lasers (e.g., the tunable nanosecond pulsed OPO systems) can routinely generate output frequencies of $\omega_1$ and $\omega_2$ that match the required vibrational modes for a variety of Raman active molecules. Moreover, it is easy to envisage that the frequency range of thus obtained CARS signals is actually controllable by choosing any conveniently available frequencies $\omega_0$ of the "common" laser source at will, because the virtual states (indicated by four dashed-dotted lines in Fig. 1) involved in the two CARS processes are not eigenstates of the Raman active molecules and hence could in principle be anywhere energetically.

To exercise the above scenario, we proceed to propose constructing an optical system, as shown schematically in Fig. 2. The entire system can be viewed as a modified type of Mach-Zehnder interferometer [10] that begins with a beam splitter (BS in Fig. 2) and ends with a beam recombiner (BR). We arrange this system in a nearly symmetrical fashion along the central dotted line which separates the entire optical layout into two segments, labeled (*a*) and (*b*) in Fig. 2. With the aid of the optical delay lines [Fig. 2 only shows the delay line (DL) in one of the $\omega_0$ beam paths.], we can make identical the optical path lengths of any laser beams between BS and BR for both segments. Each segment consists of two nonlinear frequency conversion processes (SFM and DFM) and one parametric process (CARS). As discussed below, such a method of construction will naturally make the two optical segments (*a*) and (*b*) correspond to the two optical pathways (*a*) and (*b*) shown in Fig. 1, respectively.

For convenience of description, we focus only on the optical segment (*a*). The nonlinear

SFM crystal (labeled SFMX(*a*) in Fig. 2) is angle-tuned to mix only one-half of the $\omega_0$ beam (due to the 50% BS) with one-half of the $\omega_1$ beam (due to the 50% BS1), producing an SFM beam of $\omega_p^{(a)}(=\omega_0+\omega_1)$. Right behind SFMX(*a*), a dichroic mirror (D2) is used for transmitting the residual $\omega_0$ and $\omega_1$ beams to a beam dump (DP1) while reflecting the produced $\omega_p^{(a)}$ beam. Such an $\omega_p^{(a)}$ beam is subsequently mixed with one-half of the $\omega_2$ beam (due to the 50% BS2) in a nonlinear DFM crystal (labeled DFMX(*a*)) that is angle-tuned to produce only the DFM beam of $\omega_S^{(a)}(=\omega_p^{(a)}-\omega_2=\omega_0+\omega_1-\omega_2)$. As a result, we have three collinearly propagating beams (i.e., the $\omega_p^{(a)}$, $\omega_S^{(a)}$, and residual $\omega_2$ beams [11]) in between DFMX(*a*) and the sample cell S(*a*). Eventually a CARS signal with frequency of $\omega_{CARS}^{(a)}(=2\omega_p^{(a)}-\omega_S^{(a)})$ is expected to be produced out of S(*a*), provided that the phase-matching condition is fulfilled. Since a collinear (instead of Boxcar) geometry is adopted in this paper, the CARS beam should propagate along the same direction as the collinearly propagating pump (probe) and Stokes beams. Because the frequency of thus obtained CARS signal is known *a priori*, which is equal to the frequency sum of the three source lasers used, we can readily pick only the desired CARS signal out of the unwanted residual beams by explicitly using a narrow bandpass filter (BPF1 in segment (*a*) of Fig. 2) specified for the CARS wavelength range.

It is obvious that the above descriptions for segment (*a*) should also apply to segment (*b*). When the two CARS beams of identical frequencies [cf. Eq. (3)] produced independently in the two optical segments are recombined (at BR in Fig. 2), they are expected to form stable interference patterns therein. Both CARS beams can be detected by a linear CCD (charge coupled device) array. Directing these beams to the same position on the CCD array but at slightly different angles is expected to generate a series of fringes [12]. As in standard interferometry, interference is observed only if the path-length difference between the two independent paths (*a*) and (*b*) is less than the coherence length. We adjust this path-length difference by sliding the $\omega_0$ beam splitter (BS in Fig. 2) along a linear track with an accompanying mirror (M3), which together form a retroreflector (indicated by the dashed rectangular in Fig. 2). It is noteworthy that here we adjust the path-length difference in $\omega_0$, despite the fact that interference is monitored between the two CARS beams.

To analyze the interference, we now consider the relative phase of the two CARS beams. As demonstrated in our previous work [13], the phase-matching conditions for nonlinear frequency mixing ensure that the respective phases of $\omega_p^{(i)}$ (*i* = *a, b*) can be explicitly written as the sum of the phases of the source lasers involved:

$$\phi_p^{(a)} = \phi_0 + \phi_1, \qquad \phi_p^{(b)} = \phi_0 + \phi_2, \tag{4}$$

and that the respective phases of $\omega_S^{(i)}$ (*i* = *a, b*) can be written as the difference of the phases of the source lasers involved:

$$\phi_S^{(a)} = \phi_p^{(a)} - \phi_2 = \phi_0 + \phi_1 - \phi_2, \qquad \phi_S^{(b)} = \phi_p^{(b)} - \phi_1 = \phi_0 + \phi_2 - \phi_1. \tag{5}$$

Similarly, the phase-matching conditions for four-wave mixing can lead to the following phase relationships for the two CARS beams:

$$\phi_{CARS}^{(a)} = 2\phi_p^{(a)} - \phi_S^{(a)} \equiv \phi_0 + \phi_1 + \phi_2, \tag{6a}$$

$$\phi_{CARS}^{(b)} = 2\phi_p^{(b)} - \phi_S^{(b)} \equiv (\phi_0 + \Delta\phi_0) + \phi_1 + \phi_2. \tag{6b}$$

Here the $\Delta\phi_0$ term accounts for an additional phase that is due to different path lengths for the $\omega_0$ beams between the beam splitter (BS in Fig. 2) and the two mixing crystals [SFMX(*a*) and SFMX(*b*)]. Although the individual phases of the three source lasers ($\omega_0$, $\omega_1$, and $\omega_2$) are totally uncorrelated because they all are stand-alone systems, the relative phase between the two CARS beams is fixed, since

$$\Delta\phi_{\text{CARS}} \equiv \phi^{(b)}_{\text{CARS}} - \phi^{(a)}_{\text{CARS}} = [(\phi_0 + \Delta\phi_0) + \phi_1 + \phi_2] - [\phi_0 + \phi_1 + \phi_2] = \Delta\phi_0. \tag{7}$$

Because the phase coherence between the two CARS beams is automatically assured by the above optical implementation, we have demonstrated herein that two degenerate CARS processes independently generated from two distinct Raman active media can be phase-locked, a feat that would be otherwise difficult to attain and that has, to the best of our knowledge, not yet been achieved prior to this paper.

Remarkably, thus established phase coherence between the two independent CARS processes will not be destroyed by random phase fluctuations in any of the source lasers, simply because both processes depend on the same light sources [recall Eqs. (3) and (6)], thereby making any phase fluctuations in them automatically cancel. In addition, our method of construction allows for control of the CARS interference by variation of $\Delta\phi_0$, the phase difference in the "common" laser source of frequency $\omega_0$. To change the phase difference between the two pathways [(*a*) and (*b*) in Fig. 1] at will, we can insert a rotatable glass phase plate ($\Phi$P in Fig. 2) in one of the $\omega_0$ beam paths. One would expect alternating movement (i.e., phase modulation) of the CARS fringe pattern on the CCD array when rotating the glass plate. Consequently, the relative phase difference is *not only measurable but controllable* by explicit adjustment of a phase-dependent property (here the optical thickness of a simple glass plate) known in advance, rather than inferred from other sophisticated indirect phase modulation measurements (see, e.g., [14]).

Similar to our previous work [13] in which an "interrupted" Mach-Zehnder-type interferometer was constructed towards the goal of coherently controlling chemical reactions [15] by interfering two phase-locked two-photon processes, the unique interferometer described in this paper also serves as a *phase-transport interferometer* because the phase characteristics of an input laser beam ($\omega_0$) are faithfully transferred (in a controllable manner) to two independently produced CARS beams of identical frequencies. Such a phase-transport interferometer may find potential applications in the field of quantum information processing and transfer that makes use of phase coherence properties. For instance, one can "encode" the phase information ($\Delta\phi_0$) in one optical arm of the common input source, and then "decode" (retrieve) it with high fidelity from the CARS interference pattern in a remote detector. We expect also that this interferometer can be employed to evaluate the phase coherence of CARS processes, e.g., to see how the phase modulation depth (contrast of the fringe pattern) varies when certain external conditions (e.g., temperature variations in the two Raman active samples) are changed and how the inherent non-resonant CARS background affects the observed interference. Obviously, to fully answer these interesting questions requires further experimental investigations. Based on our previous successful experience [13], we believe that such experiments (which are under way in our laboratory) using the elegant scheme proposed in this paper are feasible.

In summary, we have presented a unique approach towards phase locking of two independent degenerate CARS processes.   An interesting type of Mach-Zehnder interferometer is elaborately designed towards this goal, which itself represents an extension of the well known Young's double-slit experiment to the condensed-phase all-optical implementation of interfering two independently generated parametric processes in a phase-coherence-controllable manner.

The author is indebted to Professor Mark Keil (Ben-Gurion University, Israel) and Professor Moshe Shapiro (The University of British Columbia, Canada) for many stimulating discussions, and also thanks support from the National Natural Science Foundation of China (Grant No. 20873133), the Ministry of Science and Technology of China (Grant No. 2007CB815203), and the University of Science and Technology of China.

**REFERENCES AND NOTES**

[1] D. C. Hanna, M. A. Yuratich, and D. Cotter, *Nonlinear Optics of Free Atoms and Molecules* (Springer-Verlag, 1979).

[2] Y. R. Shen, *Principles of Nonlinear Optics* (Wiley, 1984).

[3] M. O. Scully and M. S. Zubairy, *Quantum Optics* (Cambridge U. Press, 1997).

[4] M. D. Maker and R. W. Terhune, Phys. Rev. **137**, A801 (1965).

[5] R. F. Begley, A. B. Harvey, and R. L. Byer, Appl. Phys. Lett. **25**, 387 (1974).

[6] W. M. Tolles, J. W. Nibler, J. R. McDonald, and A. B. Harvey, Appl. Spectrosc. **31**, 253 (1977).

[7] A. M. Zheltikov, J. Raman Spectrosc. **31**, 653 (2000).

[8] W. Lauterborn, T. Kurz, and M. Wiesenfeldt, *Coherent Optics: Fundamentals and Applications* (Springer-Verlag, 1993).

[9] See, e.g., C. Vinegoni, J. S. Bredfeldt, D. L. Marks, and S. A. Boppart, Opt. Express, **12**, 331 (2004); C. L. Evans, E. O. Potma, and X. S. Xie, Opt. Lett. **29**, 2923 (2004), and references therein.

[10] P. I. Hariharan, *Basics of Interferometry* (Academic, 1992).

[11] It may also contain a small amount of residual $\omega_0$ and $\omega_1$ beams that results from the non-ideal properties of the dichroic mirror (D2 in Fig. 2), which however will not affect the CARS beam later produced.

[12] J. H. Yi, S. H. Kim, and Y. K. Kwak, Meas. Sci. Technol. 11, 1352 (2000).

[13] Q. Zhang, M. Keil, and M. Shapiro, J. Opt. Soc. Am. B **20**, 2255 (2003).

[14] R. J. Gordon, S. P. Lu, S. M. Park, K. Trentelman, Y. Xie, L. Zhu, A. Kumar, and W. J. Meath, J. Chem. Phys. **98**, 9481 (1993).

[15] M. Shapiro and P. Brumer, *Principles of the Quantum Control of Molecular Processes* (Wiley, 2003).

**FIGURE CAPTIONS**

FIG. 1: (Color online) Schematic diagram for two independent degenerate CARS processes which are phase-locked by optical procedures described in the text.

FIG. 2: (Color online) Schematic diagram of the optical layout. BS, beam splitter; BR, beam recombiner; D, dichroic mirror; M, high reflecting mirror; $\Phi P$, glass phase plate; DL, delay line; SFMX, sum-frequency-mixing crystal; DFMX, difference-frequency-mixing crystal; S, Raman active sample; BPF, bandpass filter; DP, laser beam dump; CCD, charge-coupled-device linear-array detector.

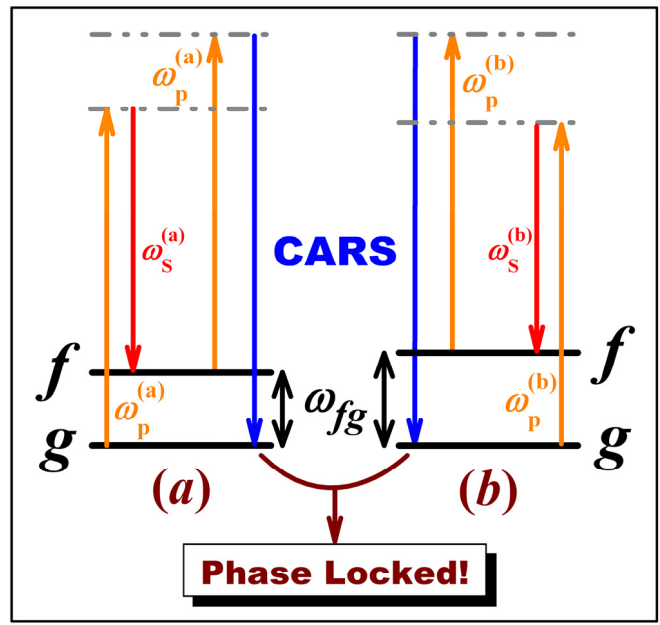

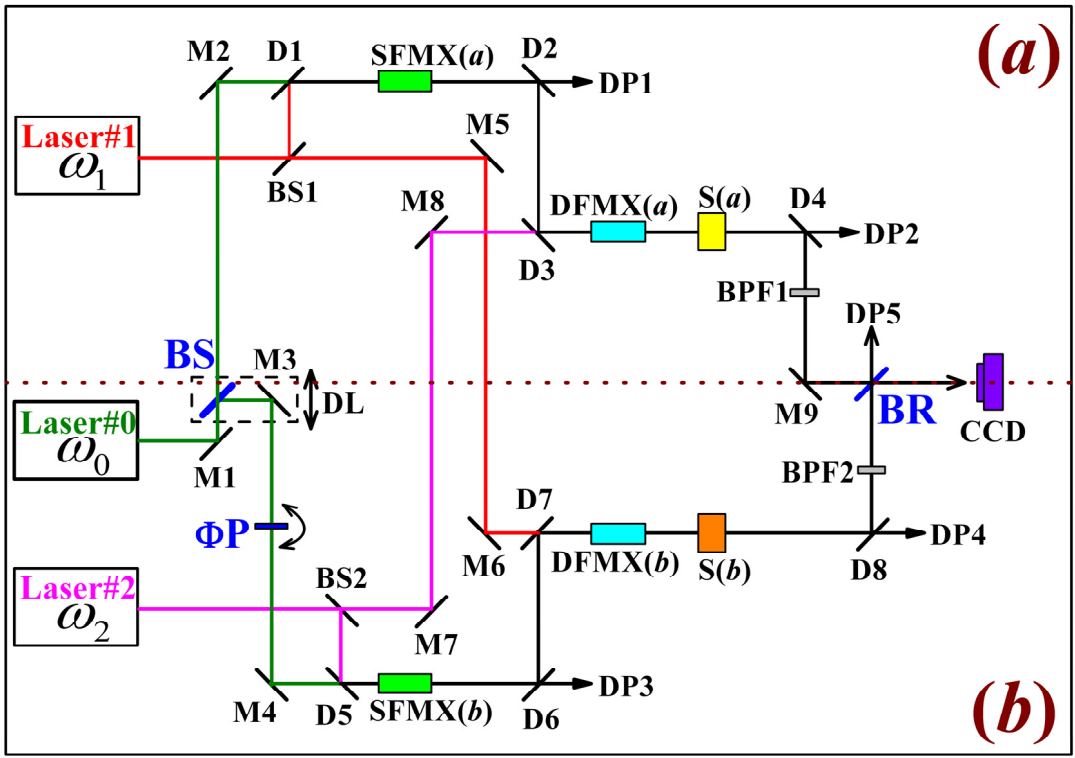